\documentclass{aa}

\usepackage{txfonts}
\usepackage{graphicx}
\usepackage{natbib}

\begin{document}

\title{X-ray optical depth diagnostics of T Tauri accretion shocks}

\author{C.~Argiroffi\inst{1,2} \and A.~Maggio\inst{2} \and G.~Peres\inst{1,2} \and J.~J.~Drake\inst{3} \and J.~L\'opez-Santiago\inst{4} \and S.~Sciortino\inst{2} \and B.~Stelzer\inst{2}}
\offprints{C.~Argiroffi, {\email argi@astropa.unipa.it}}
\institute{Dipartimento di Scienze Fisiche ed Astronomiche, Sezione di Astronomia, Universit\`a di Palermo, Piazza del Parlamento 1, 90134 Palermo, Italy, \email{argi@astropa.unipa.it, peres@astropa.unipa.it} \and INAF - Osservatorio Astronomico di Palermo, Piazza del Parlamento 1, 90134 Palermo, Italy, \email{maggio@astropa.inaf.it, sciorti@astropa.inaf.it, stelzer@astropa.inaf.it} \and Harvard-Smithsonian Center for Astrophysics, 60 Garden Street, Cambridge, MA 02138, USA \email{jdrake@cfa.harvard.edu} \and Departamento de Astrof\'isica y Ciencias de la Atm\'osfera, Facultad de Ciencias F\'isicas, Universidad Complutense de Madrid, 28040 Madrid, Spain, \email{jls@astrax.fis.ucm.es}
}
\date{Received 1 July 2009 / Accepted 25 August 2009}

\titlerunning{X-ray Optical Depth in T Tauri Stars}
\authorrunning{C.~Argiroffi et al.}

\abstract
{In classical T~Tauri stars, X-rays are produced by two plasma components: a hot low-density plasma, with frequent flaring activity, and a high-density lower temperature plasma. The former is coronal plasma related to the stellar magnetic activity. The latter component, never observed in non-accreting stars, could be plasma heated by the shock formed by the accretion process. However its nature is still being debated.}
{Our aim is to probe the soft X-ray emission from the high-density plasma component in classical T~Tauri stars to check whether this is plasma heated in the accretion shock or whether it is coronal plasma.}
{High-resolution X-ray spectroscopy allows us to measure individual line fluxes. We analyze X-ray spectra of the classical T~Tauri stars MP~Muscae and TW~Hydrae. Our aim is to evaluate line ratios to search for optical depth effects, which are expected in the accretion-driven scenario. We also derive the plasma emission measure distributions $EMD$, to investigate whether and how the $EMD$ of accreting and non accreting young stars differ. The results are compared to those obtained for the non-accreting weak-line T~Tauri star TWA~5.}
{We find evidence of resonance scattering in the strongest lines of MP~Mus, supporting the idea that soft X-rays are produced by plasma heated in the accretion shock. We also find that the $EMD$ of MP~Mus has two peaks: a cool peak at temperatures expected for plasma heated in the accretion shock, and a hot peak typical of coronal plasma. The shape of the $EMD$ of MP~Mus appears to be the superposition of the $EMD$ of a pure coronal source, like TWA~5, and an $EMD$ alike that of TW~Hydrae, which is instead dominated by shock-heated plasma.}
{}


\maketitle

\section{Introduction}

Classical T~Tauri stars (CTTS) are young low-mass stars, still surrounded by a circumstellar disk from which they accrete material. According to a widely accepted model, they have a strong magnetic field that regulates the accretion process disrupting the circumstellar disk, loading material of the inner part of the disk, and guiding it in a free fall along its flux tubes toward the central star \citep{UchidaShibata1984,BertoutBasri1988,Koenigl1991}.

A characteristic feature of young stars is strong X-ray emission that is traditionally ascribed to magnetic activity in their coronae. Similar to their more evolved siblings, the diskless weak-line T~Tauri stars (WTTS), CTTS display high X-ray luminosities and frequent flaring activity. The typical temperature of their coronal plasma is $\sim10-20$\,MK, or even higher during strong flares \citep[e.g. $50-100$\,MK,][]{GetmanFeigelson2008}.

From a theoretical point of view, the accretion process can also produce significant X-ray emission on CTTS.  Material accreting from the circumstellar disk reaches velocities of $\sim300-500\,{\rm km\,s^{-1}}$. A shock forms at the base of the accretion column because of the impact with the stellar atmosphere. This shock heats up the accreting material to a maximum temperature $T_{\rm max}=3 \mu m_{\rm H} v_{0}^2  / ( 16 k )$, where $v_{0}$ is the infall velocity. Because of the high pre-shock velocity, the infalling material reaches temperatures of a few MK, and hence it emits X-rays. Typical values of mass accretion rate for CTTS indicate that the accretion-driven X-ray luminosity should be comparable to the coronal one \citep{Gullbring1994}. Considering typical inferred stream cross-sectional
area \citep[$\la5\%$ of the stellar surface, e.g. ][]{CalvetGullbring1998}, velocity ($\sim300-500\,{\rm km\,s^{-1}}$), and mass accretion rate \citep[$\sim10^{-9}-10^{-7}\,{\rm M_{\sun}\,yr^{-1}}$, e.g.][]{GullbringHartmann1998}, it can be inferred that the plasma heated in the accretion shock should have densities $n_{\rm e}\ga10^{11}\,{\rm cm^{-3}}$, i.e. at least one order of magnitude higher than coronal plasma density. Hence, in principle, the accretion process can produce plasma with: high $L_{\rm X}$, high density, and temperatures of a few MK.

To summarize, X-ray emission from CTTS can originate from two different plasma components: plasma heated in the accretion shock and coronal plasma. The former, because of its lower temperatures, should dominate the softer X-ray band \citep[e.g. $E\le1$\,keV in the case of the CTTS TW~Hya, ][]{GuntherSchmitt2007}. While the harder X-ray emission, $E\ge1$\,keV, should be produced almost entirely by coronal plasma.

Recently, high-resolution X-ray spectra of a few CTTS enabled measurement of individual emission lines sensitive to plasma density (i.e. He-like triplets), and hence searches for evidence of accretion-driven X-rays. The density of the plasma at $T\sim2-4$\,MK can be inferred from the \ion{O}{vii} and \ion{Ne}{ix} triplet lines (at $E\approx0.6$ and 0.9\,keV, respectively). All but one of the CTTS for which the \ion{O}{vii} triplet lines were detected showed cool plasma with high density, $n_{\rm e}>10^{11}\,{\rm cm^{-3}}$ \citep{KastnerHuenemoerder2002,StelzerSchmitt2004,SchmittRobrade2005,GuntherLiefke2006,ArgiroffiMaggio2007,GuedelSkinner2007,RobradeSchmit2007}. In contrast, the cool quiescent plasma of active stellar coronae is always dominated by low densities \citep[$n_{\rm e}\la10^{10}\,{\rm cm^{-3}}$,][]{NessSchmitt2002,TestaDrake2004a}. This basic difference suggests that the high-density cool plasma in CTTS is not coronal plasma but plasma heated in  accretion shocks. One complication to this argument is that mass accretion rates derived from assuming a very high efficiency of conversion of accretion energy into X-rays tend to be an order of magnitude or so lower than rates derived using other methods \citep[e.g.][]{Drake2005,SchmittRobrade2005,GuntherSchmitt2007}.

The idea of accretion-driven X-rays from CTTS is superficially supported by a soft X-ray excess found in high-resolution X-ray spectra of CTTS with respect to similar spectra of WTTS by \citet{TelleschiGuedel2007} and \citet{GuedelTelleschi2007}. However, \citet{GuedelSkinner2007} and \citet{GuedelTelleschi2007} noted that this soft X-ray excess is significantly lower than that predicted by simple models of X-ray emission from accretion shocks. Moreover, the soft excess scales with total stellar X-ray luminosity, and hence is related at least partially in some way with the stellar magnetic activity. \citet{GuedelSkinner2007} and \citet{GuedelTelleschi2007} suggested that the CTTS soft X-rays could be produced by infalling material loaded into coronal structures.

The properties of the X-ray emitting plasma in CTTS and in WTTS have also been investigated using CCD X-ray spectra of large stellar samples. These studies, however,  commonly covered the $0.5-8.0$\,keV energy band in which the coronal component dominates.  The main results are that CTTS are on average less luminous in the X-ray band than WTTS \citep[e.g.][]{FlaccomioMicela2003}, and that X-ray emitting plasma of CTTS is on the average hotter than that of WTTS \citep{NeuhaeuserSterzik1995,PreibischKim2005}. CTTS and WTTS therefore do have different coronal characteristics, suggesting that the accretion process can affect coronal properties to some extent.

Numerical simulations have confirmed that the accretion process can produce significant X-rays: \citet{GuntherSchmitt2007} derived stationary 1-D models of the shock in an accretion column; \citet{SaccoArgiroffi2008} improved those results by performing 1-D hydrodynamical (HD) simulations of the accretion shock, including the stellar atmosphere and taking into account time variability. Assuming optically thin emission, \citet{SaccoArgiroffi2008} showed that, even for low accretion rates, the amount of X-rays produced in the accretion shock is comparable to the typical X-ray luminosity of CTTS ($L_{\rm X}\sim10^{30}{\rm erg\,s^{-1}}$ for $\dot{M}\sim10^{-10}\,{\rm M_{\odot}\,yr^{-1}}$).

Several aspects of the nature of the high-density cool plasma component observed in CTTS are still debated. In particular, definitive evidence that it is material heated in the accretion shock is still lacking.  Moreover, while the simple "photospheric burial" model of \citet{Drake2005} suggests that under some circumstances a large fraction of the shock X-rays can be absorbed and reprocessed by the photosphere, there are currently no detailed quantitative models explaining why the X-ray luminosities, predicted on the basis of 1-D HD simulation results and mass accretion rates inferred from observations at other wavelengths, are universally much higher than observed.  

Understanding the link between accretion and X-rays would also allow more accurate characterization of the coronal component of the X-ray emission from CTTS. This could help in understanding how accretion changes coronal activity, and which other parameters determine the coronal activity level in PMS stars, whose X-ray luminosity cannot simply be explained in terms of a Rossby dynamo number \citep{PreibischKim2005} as is largely the case for active main sequence stars \citep[e.g.][]{PizzolatoMaggio2003}.  

To address the above issues, we performed a detailed study of the high-resolution X-ray spectra of two nearby CTTS: TW~Hya and MP~Mus. In particular we investigated:
\begin{itemize}
\item[-] optical depth effects in their soft X-ray emission;
\item[-] the emission measure distribution {\it (EMD)} of the X-ray emitting plasma.
\end{itemize} 

Optical depth effects probe the nature of the high-density cool plasma component: we show that, if the emitting plasma is located in the accretion shock, some emission lines should have non-negligible optical depth; in contrast these lines should be optically thin if the plasma is located in coronal structures. We also investigate how the {\it EMD} can help in recognizing coronal and accretion plasma components, which should have different average temperatures. We compare the {\it EMD} of accreting and non-accreting young stars, and compare these observed {\it EMD} with that predicted on the basis of the HD shock model of \citep{SaccoArgiroffi2008}.

\section{Targets}

\begin{figure*} 
\centering
\includegraphics[width=17cm]{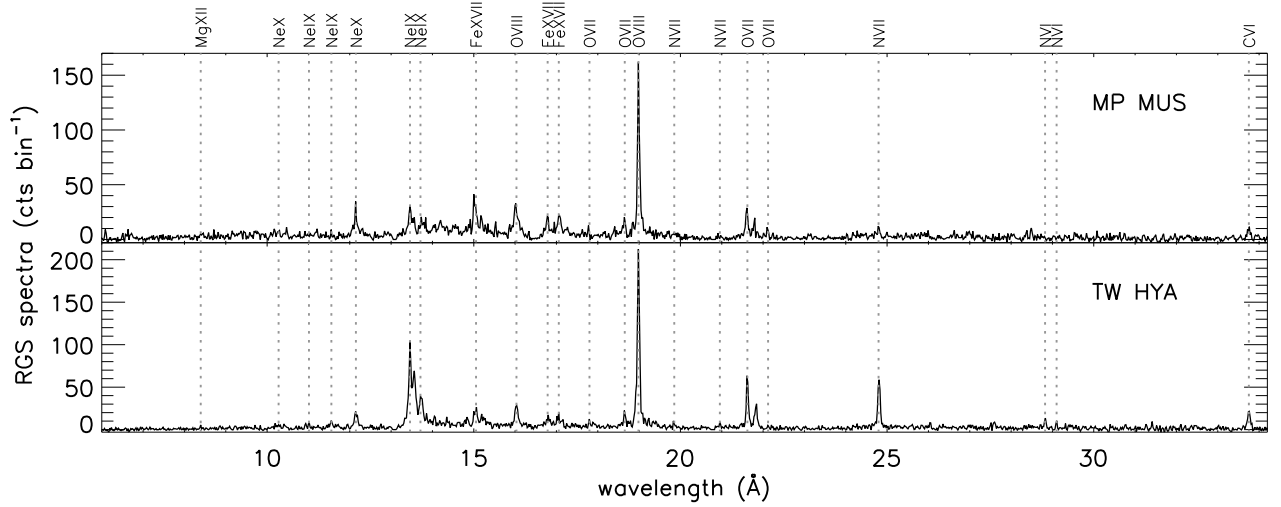} 
\caption{RGS spectra of MP~Mus and TW~Hya. Each spectrum was obtained by adding RGS1 and RGS2 spectra, rebinned by a factor 2. Labels identify some of the strongest lines.} 
\label{fig:rgsspec}
\end{figure*} 

\begin{table}[b]
\begin{center}
\caption{Log of the {\it XMM-Newton}/RGS observations of MP~Mus and TW~Hya.}
\label{tab:log}
\begin{tabular}{lcc}
\hline\hline
Instrument & Start               & Exposure \\
           & (UT)                & (ks)     \\
\hline
\multicolumn{3}{c}{MP~Mus (ObsId 0406030101)} \\
\hline
RGS1       & 2006 Aug 19 09:30:48 & 98.7    \\
RGS2       & 2006 Aug 19 09:30:56 & 98.8    \\
\hline
\multicolumn{3}{c}{TW~Hya (ObsId 0112880201 )} \\
\hline
RGS1       & 2001 Jul  9 05:51:39 & 29.0    \\
RGS2       & 2001 Jul  9 05:51:39 & 28.0    \\
\hline
\end{tabular}
\end{center}
\end{table}

\begin{table*}
\caption{Strongest RGS lines of MP~Mus and TW~Hya.}
\label{tab:lines}
\small
\begin{center}
\begin{tabular}{rlcrr@{$\;\pm\;$}lcrr@{$\;\pm\;$}lc}
\hline\hline
 & & & \multicolumn{4}{c}{MP~Mus} & \multicolumn{4}{c}{TW~Hya} \\
$\lambda_{\rm pred}^{a}$ & Ion & $\log T_{\rm max}^{b}$ & $\lambda_{\rm obs}^{a}$ & \multicolumn{2}{c}{Flux$^{c}$} & {\it EMD}$^{d}$ & $\lambda_{\rm obs}^{a}$ & \multicolumn{2}{c}{Flux$^{c}$} & 
{\it EMD}$^{d}$ \\\hline
 8.42 &                                            \ion{Mg}{xii} \ion{Mg}{xii} &  7.00 & $\cdots$ & \multicolumn{2}{c}{$\cdots$} &          &        8.39 &         4.1 &         4.1 &  $\ast$  \\
10.24 &                                                \ion{Ne}{x} \ion{Ne}{x} &  6.80 & $\cdots$ & \multicolumn{2}{c}{$\cdots$} &          &       10.28 &         7.6 &         5.2 &  $\ast$  \\
11.00 &                                                           \ion{Ne}{ix} &  6.60 & $\cdots$ & \multicolumn{2}{c}{$\cdots$} &          &       11.01 &        10.8 &         7.2 &  $\ast$  \\
11.54 &                                                           \ion{Ne}{ix} &  6.60 & $\cdots$ & \multicolumn{2}{c}{$\cdots$} &          &       11.55 &        23.6 &        16.7 &  $\ast$  \\
12.13 &                                 \ion{Ne}{x} \ion{Ne}{x} \ion{Fe}{xvii} &  6.80 &       12.14 &        22.2 &         3.4 &  $\ast$  &       12.15 &        58.6 &         9.4 &  $\ast$  \\
12.28 &                                           \ion{Fe}{xxi} \ion{Fe}{xvii} &  7.00 &       12.30 &         5.5 &         2.6 &  $\ast$  & $\cdots$ & \multicolumn{2}{c}{$\cdots$} &          \\
12.85 &                    \ion{Fe}{xx} \ion{Fe}{xx} \ion{Fe}{xx} \ion{Fe}{xx} &  7.00 &       12.86 &         3.4 &         2.3 &  $\ast$  & $\cdots$ & \multicolumn{2}{c}{$\cdots$} &          \\
13.45 &                                             \ion{Ne}{ix} \ion{Fe}{xix} &  6.60 &       13.46 &        21.1 &         3.7 &  $\ast$  &       13.46 &       244.7 &        18.5 &  $\ast$  \\
13.55 &                                             \ion{Ne}{ix} \ion{Fe}{xix} &  6.60 &       13.55 &         8.6 &         3.1 &          &       13.56 &       162.4 &        16.4 &          \\
13.70 &                                                           \ion{Ne}{ix} &  6.60 &       13.73 &         8.8 &         2.8 &          &       13.71 &        91.7 &        29.8 &          \\
14.21 &                                        \ion{Fe}{xviii} \ion{Fe}{xviii} &  6.90 &       14.19 &         3.3 &         1.7 &  $\ast$  & $\cdots$ & \multicolumn{2}{c}{$\cdots$} &          \\
14.82 &                                            \ion{O}{viii} \ion{O}{viii} &  6.50 & $\cdots$ & \multicolumn{2}{c}{$\cdots$} &          &       14.84 &        10.5 &         4.8 &  $\ast$  \\
15.01 &                                                         \ion{Fe}{xvii} &  6.70 &       15.02 &        11.8 &         2.9 &  $\ast$  &       15.05 &        26.5 &         5.7 &  $\ast$  \\
15.18 &               \ion{O}{viii} \ion{O}{viii} \ion{Fe}{xix} \ion{Fe}{xvii} &  6.50 &       15.21 &         7.6 &         2.5 &  $\ast$  &       15.21 &        23.6 &         7.7 &  $\ast$  \\
16.01 &            \ion{O}{viii} \ion{O}{viii} \ion{Fe}{xviii} \ion{Fe}{xviii} &  6.50 &       16.02 &        22.6 &         6.3 &  $\ast$  &       16.03 &        43.2 &        15.5 &  $\ast$  \\
16.78 &                                                         \ion{Fe}{xvii} &  6.70 &       16.78 &         8.8 &         1.8 &  $\ast$  &       16.79 &        13.3 &         6.8 &  $\ast$  \\
17.05 &                                          \ion{Fe}{xvii} \ion{Fe}{xvii} &  6.70 &       17.08 &        17.2 &         3.0 &  $\ast$  &       17.06 &        24.3 &         7.6 &  $\ast$  \\
17.77 &                                                           \ion{O}{vii} &  6.40 & $\cdots$ & \multicolumn{2}{c}{$\cdots$} &          &       17.80 &         5.7 &         6.6 &  $\ast$  \\
18.63 &                                                           \ion{O}{vii} &  6.30 &       18.65 &         8.7 &         2.6 &  $\ast$  &       18.65 &        21.6 &         5.8 &  $\ast$  \\
18.97 &                                            \ion{O}{viii} \ion{O}{viii} &  6.50 &       18.98 &        65.6 &         4.7 &  $\ast$  &       18.98 &       292.7 &        13.3 &  $\ast$  \\
19.83 &                                              \ion{N}{vii} \ion{N}{vii} &  6.40 & $\cdots$ & \multicolumn{2}{c}{$\cdots$} &          &       19.85 &         4.3 &         4.4 &  $\ast$  \\
20.91 &                                              \ion{N}{vii} \ion{N}{vii} &  6.30 & $\cdots$ & \multicolumn{2}{c}{$\cdots$} &          &       20.95 &        14.3 &         9.8 &  $\ast$  \\
21.60 &                                                           \ion{O}{vii} &  6.30 &       21.60 &        30.2 &         4.0 &  $\ast$  &       21.62 &       191.7 &        17.5 &  $\ast$  \\
21.80 &                                                           \ion{O}{vii} &  6.30 &       21.81 &        27.9 &         7.9 &          &       21.83 &       104.6 &        14.3 &          \\
22.10 &                                                           \ion{O}{vii} &  6.30 &       22.10 &         8.0 &         2.8 &          &       22.12 &        13.8 &         8.7 &          \\
24.78 &                                              \ion{N}{vii} \ion{N}{vii} &  6.30 &       24.80 &        10.5 &         2.8 &  $\ast$  &       24.79 &       118.2 &        13.4 &  $\ast$  \\
28.47 &                                                \ion{C}{vi} \ion{C}{vi} &  6.20 &       28.49 &         3.8 &         1.3 &  $\ast$  & $\cdots$ & \multicolumn{2}{c}{$\cdots$} &          \\
28.79 &                                                            \ion{N}{vi} &  6.20 & $\cdots$ & \multicolumn{2}{c}{$\cdots$} &          &       28.82 &        25.2 &         5.2 &  $\ast$  \\
29.08 &                                                            \ion{N}{vi} &  6.10 & $\cdots$ & \multicolumn{2}{c}{$\cdots$} &          &       29.10 &        11.5 &         5.0 &          \\
33.73 &                                                \ion{C}{vi} \ion{C}{vi} &  6.10 &       33.74 &        16.4 &         5.2 &  $\ast$  &       33.75 &        86.5 &        13.8 &  $\ast$  \\
\hline
\end{tabular}
\end{center}
$^a$~Predicted (APED database) and observed wavelengths (\AA).
$^b$~Temperature (K) of maximum emissivity.
$^c$~Observed line fluxes (${\rm 10^{-6}\,ph\,s^{-1}\,cm^{-2}}$) with uncertainties at the 68\% confidence level.
$^d$~Lines selected for the {\it EMD} reconstruction.
\normalsize
\end{table*}
\normalsize

We selected the two CTTS MP~Mus and TW~Hya for this study. These stars suffer moderate interstellar absorption ($N_{\rm H}<10^{21}\,{\rm cm^{-2}}$), and good $S/N$ spectra, gathered with {\it XMM-Newton}/RGS (high spectral resolution and large effective area in the soft X-ray band) are available.  While high-resolution X-ray spectra of TW~Hya have also been obtained by {\it Chandra}, we chose to analyze the {\it XMM-Newton}/RGS spectra so as to have more uniform data for the two stars, enabling more ready comparison of the derived results.  

TW~Hya is a $0.7\,{\rm M_{\sun}}$ CTTS, located at 56\,pc. From its membership in the eponymous TW~Hya Association (TWA), its age is estimated to be $\sim10$\,Myr \citep{KastnerZuckerman1997}. \citet{MuzerolleCalvet2000} estimated a mass accretion rate of $\sim5\times10^{-10}\,{\rm M_{\sun}\,yr^{-1}}$ based on its H$\alpha$ profile and UV flux. \citet{AlencarBatalha2002} and \citet{BatalhaBatalha2002} derived a higher value, $1-5\times10^{-9}\,{\rm M_{\sun}\,yr^{-1}}$, based on UV data, in agreement with the mass accretion rate inferred from the H$\alpha$ line width by \citet{JayawardhanaCoffey2006}. The X-ray emission of TW~Hya shows clear evidence of high-density plasma ($n_{\rm e}\sim 10^{13}\,{\rm cm^{-3}}$) at low temperatures \citep{KastnerHuenemoerder2002,StelzerSchmitt2004}.

MP~Mus is a CTTS with spectral type K1IVe. \citet{MamajekMeyer2002} identified it as a member of the Lower Centaurus Crux (LCC) Association, and determined a distance of 86\,pc using the moving cluster method and the LCC convergent point. They inferred for MP~Mus a mass of $1.1-1.2\,{\rm M_{\sun}}$ from photometry and isochrone fitting to different evolutionary tracks. Recently \citet{TorresQuast2008} suggested that MP~Mus is likely a member of the younger $\epsilon$~Cha Association. The authors, using their convergent method iteratively with their list of candidates of the association \citep{TorresQuast2006}, determined a distance to the star of 103\,pc. Note that both these distance estimates, 86 and 103\,pc, depend on correct identification of the parent stellar association.  In this work we assume a distance for MP~Mus of 86\,pc. Adopting the longer distance would lead to higher X-ray fluxes by only 30\%\ or so and would be of no consequence for the purpose of this study. Like TW Hya, MP~Mus also exhibits evidence of cool plasma at high density \citep[$n_{\rm e}\approx 5\times10^{11}\,{\rm cm^{-3}}$,][]{ArgiroffiMaggio2007}. While it is certain that MP~Mus is a CTTS \citep[H$\alpha$ equivalent width is $\sim47$\,\AA,][]{GregorioHetemLepine1992}, its mass accretion rate is not known. The accretion rate is fundamental for studying the relation between accretion and soft X-ray emission, and we evaluate it in Sect.~\ref{optobs} using an optical spectrum of MP~Mus taken with FEROS at La Silla Observatory.

We compare the results obtained for TW~Hya and MP~Mus with those obtained for TWA~5, a member of the TWA with similar age to TW~Hya. TWA~5 is a quadruple system with three similar components of $0.5\,{\rm M_{\odot}}$ and a brown dwarf. H$\alpha$ equivalent width \citep[$\sim10$\,\AA,][]{GregorioHetemLepine1992} indicates that TWA~5 is a WTTS, even if some marginal accretion is still present at least in one of the components \citep{MohantyJayawardhana2003}. We selected TWA~5 for the comparison because: 1) its {\it EMD} was derived from an {\it XMM-Newton}/RGS observation \citep{ArgiroffiMaggio2005} with the same method we adopted in this work for MP~Mus and TW~Hya; 2) low plasma density was inferred from the \ion{O}{vii} triplet \citep{ArgiroffiMaggio2005}, suggesting that its X-ray emission is entirely produced by the stellar corona.

\section{RGS data analysis}

TW~Hya was observed on July 2001 for 30\,ks, and MP~Mus on August 2006 for 110\,ks. Details of the data processing can be found in \citet{StelzerSchmitt2004} and \citet{ArgiroffiMaggio2007}, respectively. In Table~\ref{tab:log} we report the log of the {\it XMM-Newton}/RGS observations of the two stars. For MP~Mus we discarded time segments affected by high background count rates, obtaining a net exposure of $\sim100$\,ks over the selected good time intervals.

Background-subtracted spectra of MP~Mus and TW~Hya contain $\sim5800$ and $\sim6500$ net counts, respectively. To increase the $S/N$ ratio of the data we added the RGS1 and RGS2 spectra and rebinned by a factor 2. The resulting RGS spectra are shown in Fig.~\ref{fig:rgsspec}. 

We based the spectral analysis (line identification, line emissivity functions $G(T)$, line oscillator strengths $f$) on the Astrophysical Plasma Emission Database \citep[APED V1.3,][]{SmithBrickhouse2001}, as implemented in the PINTofALE V2.0 software package \citep{KashyapDrake2000}. Spectral line fluxes were measured within PINTofALE by fitting the observed lines with a Lorentzian profiles. The results are listed in Table~\ref{tab:lines}.

\subsection{EMD reconstruction}
\label{emd}

Having defined a temperature grid, $T_{i}$, with binsize $\Delta T$, the {\it EMD} of the X-ray emitting plasma is computed as the amount of emission measure, $n_{\rm e}\,n_{\rm H}\,\Delta V$, of plasma with temperature ranging between $T_{i}-\Delta T /2$ and $T_{i}+\Delta T /2$.  We derived the {\it EMD} from the line flux measurements using the Markov-Chain Monte Carlo (MCMC) method of \citet{KashyapDrake1998}. We considered all the measured lines whose intensity depends only on the plasma temperature, and not on the plasma density: hence, we discarded the intercombination and forbidden lines of the \ion{O}{vii} and \ion{Ne}{ix} He-like triplets. The lines used for the {\it EMD} reconstruction are marked in Table~\ref{tab:lines}.

We reconstructed the plasma {\it EMD} over a logarithmic temperature grid ranging between $\log T {\rm (K)}=6.0$ and 7.2, with a bin size of 0.2. This choice was guided by the set of formation temperatures of the measured lines (see $\log T_{\rm max}$ in Table~\ref{tab:lines}). This range includes the expected temperature of plasma heated in an accretion shock. On the other hand, coronal plasma may also have temperatures higher than $\log T =7.2$, so the derived {\it EMD} does not take very hot coronal components into account.

The derived {\it EMD} for MP~Mus and TW~Hya are reported in Table~\ref{tab:emd}. Simultaneously with the {\it EMD} reconstruction, we derived also the relative abundances of elements represented by the selected lines. The absolute abundances, also reported in Table~\ref{tab:emd}, were fixed for both stars by matching the predicted and observed spectra.

\begin{table}[t]
\renewcommand{\baselinestretch}{1.3}
\caption{EMD and abundances of MP~Mus and TW~Hya.}
\label{tab:emd}
\normalsize
\begin{center}
\begin{tabular}{ccc}
\hline\hline
 & MP~Mus & TW~Hya \\
\hline
$\log T$~(K) & $\log\,${\it EMD}\,$({\rm cm^{-3}})$ & $\log\,${\it EMD}\,$({\rm cm^{-3}})$ \\
\hline
 $  6.00$ & $ 51.76^{+  0.76}_{-  0.19}$ & $ 51.79^{+  0.25}_{-  0.08}$ \\
 $  6.20$ & $ 52.24^{+  0.42}_{-  0.27}$ & $ 51.74^{+  0.25}_{-  0.05}$ \\
 $  6.40$ & $ 53.03^{+  0.12}_{-  0.42}$ & $ 52.64^{+  0.04}_{-  0.10}$ \\
 $  6.60$ & $ 53.00^{+  0.05}_{-  0.20}$ & $ 52.29^{+  0.05}_{-  0.06}$ \\
 $  6.80$ & $ 52.76^{+  0.16}_{-  0.08}$ & $ 51.41^{+  0.12}_{-  0.09}$ \\
 $  7.00$ & $ 52.89^{+  0.17}_{-  0.20}$ & $ 51.63^{+  0.08}_{-  0.13}$ \\
 $  7.20$ & $ 53.13^{+  0.11}_{-  0.66}$ & $ 51.69^{+  0.08}_{-  0.21}$ \\
\hline
 Elem. & $A_{X}/A_{X\,\sun}$ & $A_{X}/A_{X\,\sun}$ \\
\hline
  C & $  0.25^{+  0.10}_{-  0.09}$ & $  0.70^{+  0.15}_{-  0.17}$ \\
  N & $  0.28^{+  0.14}_{-  0.10}$ & $  2.23^{+  0.45}_{-  0.44}$ \\
  O & $  0.18^{+  0.09}_{-  0.03}$ & $  0.96^{+  0.19}_{-  0.10}$ \\
 Ne & $  0.46^{+  0.14}_{-  0.06}$ & $  7.66^{+  1.30}_{-  0.94}$ \\
 Mg &                  $={\rm Fe}$ & $  2.30^{+  1.84}_{-  2.14}$ \\
 Fe & $  0.09^{+  0.00}_{-  0.00}$ & $  0.43^{+  0.00}_{-  0.00}$ \\
\hline
\end{tabular}
\end{center}
\renewcommand{\baselinestretch}{1.0}
Abundances are in solar units \citep{AsplundGrevesse2005}. EMD values have been derived assuming that MP~Mus and TW~Hya are located at 86 and 56\,pc, respectively.
\normalsize
\end{table}
\normalsize

\subsection{Optical depth effects}
\label{optdepth}

Assuming that the X-ray emitting plasma has a geometrical depth $l$ (in cm) along the line of sight, free electrons with density $n_{\rm e}$ and temperature $T_{\rm e}$ (in K), then the optical depth of an emission line is given by \citep{Acton1978}:

\begin{equation}
\tau = 1.16\times10^{-14}\,
\lambda\,f\,
\left( \frac{m_{Z}}{T_{\rm e}} \right)^{1/2} \,
\frac{n_{Z, i}}{n_{Z}}\,
\frac{n_{Z}}{n_{\rm H}}\,
\frac{n_{\rm H}}{n_{\rm e}}\,
n_{\rm e}\,l
\label{eq:tau}
\end{equation}

\noindent
where $\lambda$ is the line wavelength (in \AA), $f$ the line oscillator strength, $m_{Z}$ is the atomic mass of the element (in amu), $n_{Z, i}$ the density of element $Z$ with ionization level $i$, $n_{Z}$ the density of element $Z$, $n_{\rm H}$ the hydrogen density (all the densities are in units of ${\rm cm^{-3}}$).

For increasing density $n_{\rm e}$, and/or increasing source dimension $l$, $\tau$ increases. A non-negligible optical depth ($\tau\sim 1$), because of resonance scattering, first occurs in lines with large oscillator strengths $f$ and ground state lower levels.  Photons produced in these transitions are absorbed again by ions of the same species in the ground state.  These photons are then re-emitted in different, random directions. When the optical depth is of the order of unity, resonance scattering only affects the emergent spectrum if the source does not have a spherical symmetry.

In the scenario of accretion-driven X-ray emission, the material heated in the accretion shock is entirely located in a small compact volume at the footpoint of the accretion stream. The high-density cool plasma component observed in CTTS has density of $10^{11}-10^{13}\,{\rm cm^{-3}}$ and linear dimension of $l\sim10^{9}-10^{10}$\,cm \citep[inferred from measured values of $n_{\rm e}$ and {\it EM},][]{KastnerHuenemoerder2002,StelzerSchmitt2004,ArgiroffiMaggio2007,RobradeSchmit2007}. Considering these values for $n_{\rm e}$ and $l$, the optical depth of the strongest emission lines produced by this shock-heated plasma ($T\sim1-3$\,MK) should be non-negligible. As an example, the line center optical depth, $\tau$, of the \ion{O}{vii} resonance line at 21.60\,\AA~ is $\tau\sim10$ for a density $n_{\rm e}=1\times10^{11}\,{\rm cm^{-3}}$, $n_{Z,i}/n_{Z}=0.5$ (corresponding
to a temperature $T=2$\,MK), $n_{Z}/n_{\rm H}=8.5\times10^{-4}$, $n_{\rm H}/n_{\rm e}=0.83$, $m_{Z}=16.0$\,amu, and $l=10^{9}$\,cm.

On the other hand, if this high-density plasma component observed in CTTS is contained in many separate coronal structures, the average length $l$ covered by photons before escaping from the plasma should be significantly smaller, and hence the optical depth negligible. In fact, extensive investigation of coronal spectra for optical depth effects 
has shown coronae to be effectively optically-thin in almost all cases \citep[e.g.][]{NessSchmitt2003,TestaDrake2004b,TestaDrake2007}, with only a few notable exceptions (see below).

Optical depth effects in a given line can be investigated by considering, as a reference, another line produced by the same element in the same ionization stage, but with a smaller oscillator strength \citep{NessSchmitt2003,TestaDrake2004b,MatrangaMathioudakis2005,TestaDrake2007}. In the optically-thin case, the ratio between two such lines is dictated purely by atomic parameters and has only a weak dependence on the plasma temperature. Instead, when the gas is not optically-thin, the line with larger oscillator strength can be quenched relative to the weaker line, causing a discrepancy between the observed and predicted optically-thin ratios.

In the high-resolution X-ray spectra gathered with {\it Chandra} and {\it XMM-Newton}, the strongest emission lines with the largest oscillator strengths are the Ly$\alpha$ lines of \ion{O}{viii} and \ion{Ne}{x}, the resonance lines of \ion{O}{vii} and \ion{Ne}{ix}, and the \ion{Fe}{xvii} line at 15.01\,\AA. Considering {\it Chandra}/HETGS data, \citet{TestaDrake2004b} and \citet{TestaDrake2007} found significant resonance scattering in the Ly$\alpha$/Ly$\beta$ line ratios of \ion{O}{viii} and \ion{Ne}{x} of the active stars II~Peg and IM~Peg. \citet{MatrangaMathioudakis2005} and \citet{RoseMatranga2008}, analyzing RGS data of AB~Dor and EV~Lac, found that the ratio between the \ion{Fe}{xvii} lines at 15.01 and 16.78\,\AA~changes between the flaring and quiescent phases, indicating opacity in the 15.01\,\AA~line.

\begin{table}[t]
\renewcommand{\baselinestretch}{1.3}
\caption{Line ratios sensitive to optical depth.}
\label{tab:linerat}
\normalsize
\begin{center}
\begin{tabular}{lrccc}
\hline\hline
Star & $N_{\rm H}\,({\rm cm^{-2}})$ & $\frac{\ion{O}{vii}(r)}{\ion{O}{vii}(18.6)}$ & $\frac{\ion{O}{viii}(Ly\alpha)}{\ion{O}{viii}(Ly\beta)}$ & $\frac{\ion{Fe}{xvii}(15.0)}{\ion{Fe}{xvii}(16.8)}$  \\
\hline
MP~Mus       & $5\times10^{20}$   & $4.0\pm1.5$   & $3.2\pm1.0$   & $1.3\pm0.4$   \\
TW~Hya       & $2\times10^{20}$   & $9.4\pm2.8$   & $7.0\pm2.6$   & $1.9\pm1.1$   \\
TWA~5        & $3\times10^{20}$   & $8\pm4$       & $11\pm4$      & $1.4\pm0.4$   \\
\hline
Procyon      & $1.1\times10^{18}$ & $7.9\pm1.1$   & $\cdots$      & $1.2\pm0.4$   \\
AT~Mic       & $1\times10^{18}$   & $6.2\pm2.6$   & $\cdots$      & $1.8\pm0.6$   \\
$\delta$~CrB & $\cdots$           & $\cdots$      & $\cdots$      & $2.4\pm0.4$   \\
YZ~CMi       & $1.2\times10^{18}$ & $\cdots$      & $\cdots$      & $1.2\pm0.4$   \\
44~Boo       & $1\times10^{18}$   & $\cdots$      & $\cdots$      & $2.39\pm0.28$ \\
$\alpha$~Cen & $\cdots$           & $9.7\pm1.0$   & $\cdots$      & $\cdots$      \\
Capella      & $1.8\times10^{18}$ & $\cdots$      & $\cdots$      & $1.64\pm0.12$ \\
AB Dor       & $1.7\times10^{19}$ & $8.9\pm1.6$   & $\cdots$      & $1.85\pm0.10$ \\
\hline
\end{tabular}
\end{center}
\renewcommand{\baselinestretch}{1.0}
Ratios are derived from line fluxes in ${\rm ph\,s^{-1}\,cm^{-2}}$.
Active stellar measurements are taken from: \citet{RaassenMewe2002} for Procyon; \citet{RaassenMewe2003} for AT~Mic; \citet{Gondoin2005} for $\delta$~CrB; \citet{RaassenMitra-Kraev2007} for YZ~CMi ; \citet{Gondoin2004} for 44~Boo; \citet{AudardBehar2001} for Capella; \citet{SanzForcadaMaggio2003} for AB~Dor.
\normalsize
\end{table}
\normalsize

We searched for optical depth effects in the X-ray emission of MP~Mus and TW~Hya, using TWA~5 as a non-accreting comparison.  We investigated three line ratios:
\begin{itemize}
\item[-] the ratio of the \ion{O}{vii} lines at 21.60\,\AA~and 18.63\,\AA, produced by transitions to the ground level ($n=1$) from the shells $n=2$ and $n=3$ respectively; the resonance line at 21.60\,\AA~has a large oscillator strength ($f=0.69$), while the line at 18.63\,\AA~has a small oscillator strength ($f=0.15$);
\item[-] the ratio between the \ion{O}{viii} Ly$\alpha$ at 18.97 (a doublet with oscillator strengths $f=0.54$ and $f=0.27$) and the \ion{O}{viii} Ly$\beta$ at 16.01\,\AA~(a doublet with $f=0.10$ and $f=0.05$);
\item[-] the ratio between the \ion{Fe}{xvii} lines at 15.01\,\AA~(large oscillator strength, $f=2.73$) and 16.78\,\AA~(small oscillator strength, $f=0.11$).
\end{itemize}
This line set allowed us to probe the optical depth of the plasma at temperatures ranging from 2 to 5\,MK. We did not consider the \ion{Ne}{x} Ly$\alpha$ and Ly$\beta$ line ratio since, for MP~Mus, we detected only the Ly$\alpha$ line.

\begin{figure} 
\resizebox{\hsize}{!}{\includegraphics{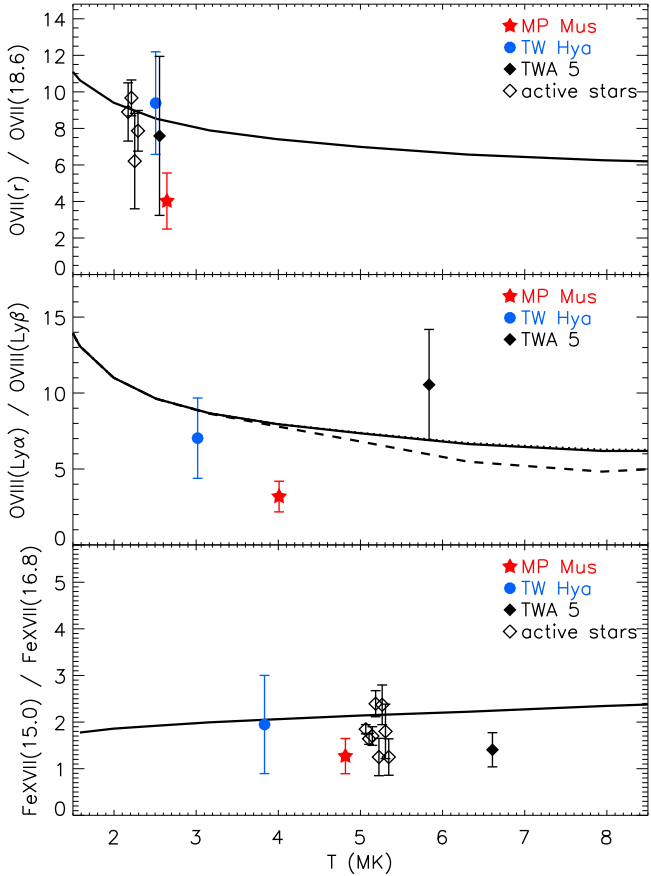}}
\caption{Line photon flux ratios sensitive to optical depth. Filled symbols mark the observed line ratios, corrected for interstellar absorption, for the three stars MP~Mus, TW~Hya, and TWA~5. Each ratio is associated with the average temperature obtained assuming the function {\it EMD}$(T)\times G(T)$, as weighting function. The temperature associated to the observed ratios for the active stars (open symbols) selected is the temperature corresponding to the maximum value of the emissivity function $G(T)$ of the 21.60 and 15.01\,\AA~lines. Predicted values are computed assuming optically thin emission from an isothermal plasma.
{\it Upper panel}: photon flux ratio of the \ion{O}{vii} lines at 21.60 and 18.63\,\AA.
{\it Medium panel}: photon flux ratio of the \ion{O}{viii} Ly$\alpha$ vs. Ly$\beta$ lines. Due to a blend with Fe lines, predicted values differ for the three stars because of the different relative abundances. The solid line indicates the ratio predicted for MP~Mus, the dotted line (barely visible below the solid line) for TW~Hya, and the dashed line for TWA~5.
{\it Lower panel}: photon flux ratio of the \ion{Fe}{xvii} lines at 15.01 and 16.78\,\AA.
}
\label{fig:fluxratio} 
\end{figure} 

The observed ratios of MP~Mus, TW~Hya, and TWA~5 are listed in Table~\ref{tab:linerat} and plotted in Fig.~\ref{fig:fluxratio}. These ratios were obtained correcting the observed fluxes (Table~\ref{tab:lines}) for the interstellar absorption, assuming the  $N_{\rm H}$ values derived from the analysis of EPIC spectra \citep{ArgiroffiMaggio2007,StelzerSchmitt2004,ArgiroffiMaggio2005}. The correction factors due to interstellar absorption range between 0.94 and 1.16. We associated each flux ratio of each star to an average temperature. That average temperature was obtained assuming, as a weighting function, the product of the {\it EMD} and emissivity functions, $G(T)$, of the line in the numerator. The predicted flux ratios as a function of plasma isothermal temperature are also illustrated in Fig.~\ref{fig:fluxratio}.

All the lines involved in the analyzed ratios are free from significant blending, except for the \ion{O}{viii} Ly$\beta$ line, located at 16.01\,\AA. This line is blended with two \ion{Fe}{xviii} lines (16.00 and 16.07\,\AA), whose contributions are small but non-negligible \citep[see, e.g.][for a detailed discussion and treatment]{TestaDrake2007}: the {\it EMD} and abundances derived for the X-ray emitting plasma of MP~Mus and TW~Hya (see Sect.~\ref{emd}) indicate that the \ion{Fe}{xviii} contribution to the observed fluxes is about $\sim20\%$ and $\sim6\%$, respectively. To take this into account, we included these contributions in the calculation of the predicted flux ratios. The predicted ratio of the \ion{O}{viii} Ly$\alpha$ and Ly$\beta$ lines is then weakly dependent on the model abundances and hence differs slightly for different stars. 

For comparison, we considered also other coronal sources for which line fluxes and $N_{\rm H}$ values, obtained from {\it XMM-Newton}/RGS data, were published. These stars and their \ion{O}{vii} and \ion{Fe}{xvii} line flux ratios, corrected for the interstellar absorption, are listed in Table~\ref{tab:linerat} and are included in Fig.~\ref{fig:fluxratio}. We did not consider the \ion{O}{viii} line ratio since its analysis requires knowledge of model abundances for each star.

\begin{figure} 
\resizebox{\hsize}{!}{\includegraphics{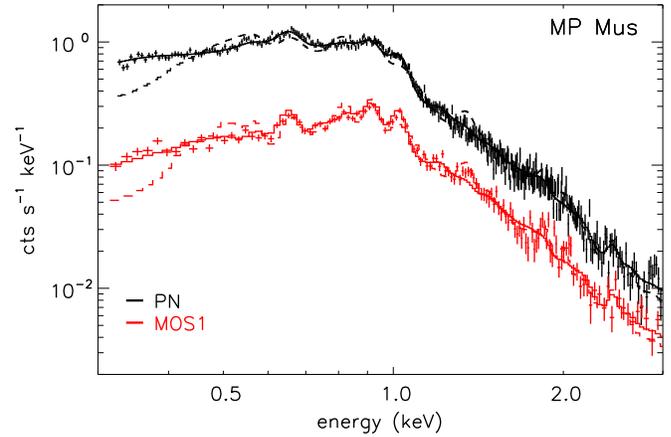}}
\caption{PN (black) and MOS1 (red) spectra of MP~Mus with superimposed the 3$-T$ model (solid line) derived by \citet{ArgiroffiMaggio2007} indicating low absorption ($N_{\rm H}=5\times10^{20}\,{\rm cm^{-2}}$), and the 3$-T$ model (dashed line) obtained by fitting the EPIC spectra with an absorbing column fixed at $5\times10^{21}\,{\rm cm^{-2}}$.}
\label{fig:mpmusepicspec} 
\end{figure} 

MP~Mus shows in all cases an observed ratio significantly lower than the predicted one, over the entire range of plausible temperature. The discrepancy between the observed and predicted ratios for MP~Mus are 2.8$\sigma$, 4.7$\sigma$, and 2.3$\sigma$ for the \ion{O}{vii}, \ion{O}{viii}, and \ion{Fe}{xvii} ratios, respectively. Conversely, the observed ratios for TW~Hya are perfectly compatible with those predicted for optically-thin emission. The \ion{O}{vii} and \ion{O}{viii} ratios for all the other stars considered are compatible with the predicted  optically-thin values. Conversely some stars, other than MP~Mus, show a \ion{Fe}{xvii} line ratio discrepant with the optically thin case. The interpretation of the \ion{Fe}{xvii} ratio is more controversial: while coronal plasma might have non-negligible optical depth in the 15.01\AA~line \citep[i.e. the Sun and AB~Dor,][]{SabaSchmelz1999,MatrangaMathioudakis2005}, \citet{NessSchmitt2003} found that several stars, in a large sample, exhibit ratios somewhat discrepant from the predicted optically-thin value, as in our case, but they concluded that the interpretation of this result in terms of opacity effects is not convincing.

Summarizing, we detected a significant intensity deficit in lines with the largest oscillator strength in the spectrum of MP~Mus: the \ion{O}{vii} resonance line at 21.60\,\AA, the \ion{O}{viii} Ly$\alpha$ line at 18.98\,\AA, and the \ion{Fe}{xvii} line at 15.01\,\AA. These lines are mainly formed by the plasma at a temperature of $2-8$\,MK.  We conclude that this emission is not optically-thin and that resonance scattering has quenched the  strongest lines. 

We explored whether the anomalous oxygen line ratios observed in MP~Mus could be explained by an absorbing column $N_{\rm H}$ higher than that assumed (that however could not explain the \ion{Fe}{xvii} line ratio). To explain the observed \ion{O}{vii} and \ion{O}{viii} line ratios for MP~Mus, a hydrogen column $\ga 5\times10^{21}\,{\rm cm^{-2}}$ is required. This value is higher, by a factor 10, than the value constrained from the analysis of the EPIC spectra by \citet{ArgiroffiMaggio2007}, and adopted for the line ratio analysis. In Fig.~\ref{fig:mpmusepicspec} we show the PN and MOS1 spectra of MP~Mus (MOS2 is omitted for clarity). We also plotted the 3$-T$ model with the inferred absorption $N_{\rm H}=(5 \pm 1.8)\times10^{20}\,{\rm cm^{-2}}$ \citep{ArgiroffiMaggio2007}, and the 3$-T$ model obtained by fitting the EPIC spectra with an absorbing column fixed at $5\times10^{21}\,{\rm cm^{-2}}$. The soft part of the EPIC spectra, which constrains the $N_{\rm H}$ value, is dominated by continuum emission. Therefore, if soft continuum and soft RGS lines, both contained in the $0.3-1.0$\,keV energy range, are produced by the same plasma component, then the observed line ratios of MP~Mus cannot be explained in terms of photoelectric absorption.  Instead if soft continuum and soft lines originated from different plasma components, affected by different absorbing columns, then the observed line ratios could be explained by photoelectric absorption, rather than opacity effects in the emitting plasma. However, different absorbing columns would indicate that the highly absorbed component (the one producing the \ion{O}{vii} and \ion{O}{viii} lines) is located under the accretion stream or buried down in the stellar atmosphere, suggesting again that this plasma component originates in the accretion shock.

\section{Optical spectra and their analysis}
\label{optobs}

To gather information on the accretion status of MP~Mus, we analyzed its optical spectrum taken from the public archive of La Silla Observatory (ESO). The observation was performed on March 18, 2005 using the echelle spectrograph FEROS at the 2.2m telescope. The spectra cover approximately 3850--9200 \AA\ with a resolving power $R = 48000$ ($\Delta\lambda \approx 0.15$ \AA, in the region of H$\alpha$, measured in the calibration lamp spectrum). This range includes all the chromospheric activity indicators, ranging from \ion{Ca}{ii} H \& K to the calcium infrared triplet, as well as the lithium line at 6708 \AA. The signal-to-noise ratio at 6550 \AA\ is $S/N \sim 75$.

For the reduction, we used the standard procedures in the IRAF\footnote{IRAF is distributed by the National Optical Observatory, which is operated by the Association of Universities for Research in Astronomy, Inc., under contract with the National Science Foundation.} package (bias subtraction, extraction of the scattered light produced by the optical system, division by a normalized flat-field, and wavelength calibration). After reduction, the spectrum of MP~Mus was normalized to the continuum order by order by fitting a polynomial function to remove the general shape from the aperture spectra. We did not perform a flux calibration since it is not necessary for our study.

\subsection{H$\alpha$ line and accretion rate}

We measured the equivalent width of the H$\alpha$ line using an IDL procedure developed by us that enables estimation of the error in the measurements using both the signal-to-noise ratio and the spectral resolution. We obtained $EW({\rm H\alpha})=-41.05\pm0.09$\,\AA, which is similar to the value reported by \citet{MamajekMeyer2002}, but lower than that measured by \citet{GregorioHetemLepine1992}, who found $EW({\rm H\alpha})=-47$\,\AA. This variation of 6\,\AA\ in observations obtained $\sim10$\,yr apart is not very high if compared e.g. with variations observed by \citet{SaccoFranciosini2008} in $\sigma$~Ori and $\lambda$~Ori. However, there is not enough time coverage to study the H$\alpha$ variability in detail. 

\begin{figure} 
\resizebox{\hsize}{!}{\includegraphics{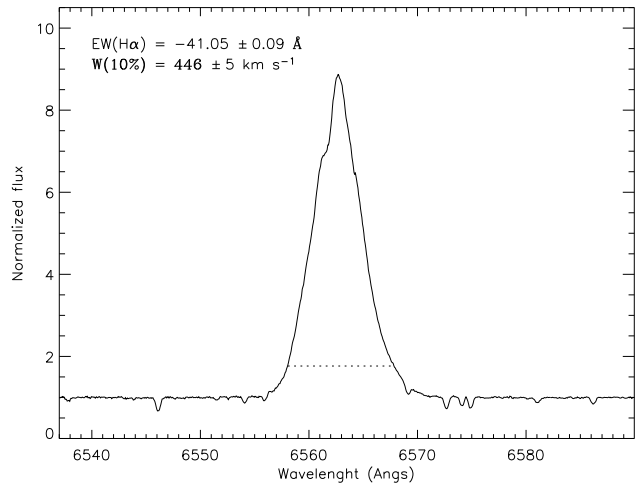}}
\caption{H$\alpha$ line profile of MP~Mus, illustrating the line width at 10\% of the maximum.}
\label{fig:halpha} 
\end{figure}

We also measured the width at 10\% of the peak of the H$\alpha$ line (see Fig.~\ref{fig:halpha}), which is indicative of the mass accretion rate for widths above $200\,{\rm km\,s^{-1}}$ \citep{NattaTesti2004}. We obtained $W({\rm H\alpha})=446\pm5$ km\,s$^{-1}$, corresponding to a mass accretion rate of $\approx3\times10^{-9}\,{\rm M_{\odot}\,yr^{-1}}$.

\section{Discussion}

\subsection{Soft X-rays from shock-heated plasma: optical depth effects}

It appears that in most CTTS, significant soft X-rays are produced by a high-density plasma component. This result has been interpreted as a strong indication that this plasma, at a temperature of a few MK, is formed in the shock at the base of the accretion stream near the stellar surface. Hence this high-density plasma component should be contained in a small volume at the base of the accretion stream, instead of being located in extended structures like coronal plasma.

We showed in Sect.~\ref{optdepth} that, considering the characteristic volumes and densities for the high-density cool plasma component of CTTS, the optical depth $\tau$ (eq.~\ref{eq:tau}) in some emission lines with large oscillator strength should be significantly larger than 1, producing detectable opacity effects.
We found clear evidence of opacity effects in all the examined line ratios of the soft X-ray spectrum of MP~Mus, produced by the high-density plasma at temperatures of a few MK. These lines are: the resonance lines produced by $n=2$ to $n=1$ shell transitions in the \ion{O}{vii} and \ion{O}{viii} ions (at $21.60$\,\AA~and 18.97\,\AA, respectively), and the \ion{Fe}{xvii} line at 15.01\,\AA. All these lines are significantly weaker than their predicted optically-thin intensities, a result which can be readily explained in terms of resonance scattering. We did not find evidence of optical depth effects in the X-ray emission from TW~Hya.

We explored whether the observed opacity could be produced by coronal plasmas. Studies on large stellar surveys showed that optical depth effects from cool coronal plasma exist but are extremely rare \citep{NessSchmitt2003,TestaDrake2007}. Quiescent plasma confined within coronal magnetic structures can hardly explain the observed non-negligible optical depths in CTTS. In fact the \citet{RosnerTucker1978} model indicates loop semi-length of $\sim10^{6}-2\times10^{7}$\,cm, for $T_{\rm max}=3$\,MK and $n_{\rm e}=5\times10^{11}-10^{13}\,{\rm cm^{-3}}$. Such scale lengths are unrealistically small if compared to photospheric and chromospheric scale heights, and not large enough to account for the observed optical depth ($\tau$ would be $\sim0.03$, in the center of the \ion{O}{vii} line at 21.60\,\AA, assuming that the loop cross section has a radius smaller by a factor 10 than the loop semi-length). In principle flaring plasma, instead of quiescent plasma, could favor opacity effects. In fact standard models suggest that plasma density increases during flares, raising the opacity. Opacity effects have been searched in flaring coronal sources \citep[e.g.][]{GuedelAudard2004}, but they have been detected only a couple of times \citep{MatrangaMathioudakis2005,RoseMatranga2008}. In those cases opacity affected \ion{Fe}{xvii} lines formed at higher temperatures than the \ion{O}{vii} and \ion{O}{viii} lines. Moreover the soft X-ray light curve of MP~Mus indicates that this emission does not originate from flaring plasma \citep{ArgiroffiMaggio2007}.

We cannot reject altogether the hypothesis that the high-density cool plasma of MP~Mus, showing opacity effects, originates from coronal plasma. However the alternative explanation, based on the hypothesis that this plasma is located in the post shock region at the base of the accretion stream, naturally explains the observed opacity effects. Therefore opacity by itself is not a prove of the nature of this plasma component, but strongly supports the shock-heated plasma hypothesis.

We find evidence of optical depth effects for MP~Mus, and not for TW~Hya, although optical depth effects were expected for both stars. Moreover the higher densities of the plasma of TW~Hya, in comparison to MP~Mus, should favor opacity effects, considering that the characteristic dimension of the plasma volume $l$ scales with $EM^{1/3}\,n_{\rm e}^{-2/3}$, and hence $\tau\propto l\,n_{\rm e}=EM^{1/3}\,n_{\rm e}^{1/3}$. A tentative explanation of the lack of opacity effects from TW~Hya could be related to the fact that resonance scattering depends on the geometry of the source and on the inclination angle under which the source is observed. In particular, the post-shock accretion hot-spot is likely to have quite different vertical and horizontal dimensions.

\begin{figure} 
\resizebox{\hsize}{!}{\includegraphics{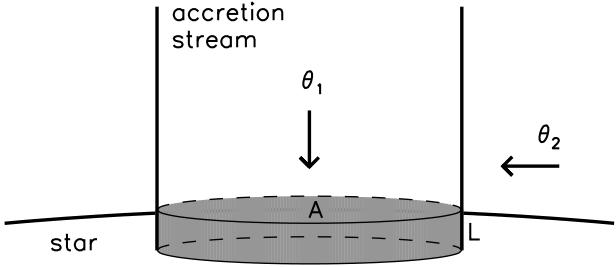}}
\caption{Schematic view of the accretion stream footpoint at the stellar surface. The gray volume indicates the post-shock region containing the high-density plasma at temperatures of a few MK.}
\label{fig:accretion} 
\end{figure}

The volume occupied by the plasma heated in the accretion shock is defined by the stream cross-section $A$ and by the thickness of the hot post-shock region $L$ (see Fig.~\ref{fig:accretion}). $L$ is mainly determined by the post-shock velocity and by the plasma cooling time, and hence by the plasma density \citep[e.g.][]{SaccoArgiroffi2008}. Estimates of both $A$ and $L$ can be made from the soft X-ray emission of CTTS: observed electron densities range from $10^{11}\,{\rm cm^{-3}}$ to $10^{13}\,{\rm cm^{-3}}$, a typical value of the $EM=n_{\rm e}^2 \times A \times L$ responsible for the soft X-ray emission is $\sim10^{53}\,{\rm cm^{-3}}$. Assuming an infall velocity of $400\,{\rm km\,s^{-1}}$ and assuming that all the shock-heated plasma is contained in one shock area, we obtain: $L\sim10^{7}-10^{9}\,{\rm cm}$ and $A\sim10^{20}-10^{22}\,{\rm cm^{2}}$. The difference of the horizontal and vertical dimensions does not depend on assuming only one shock area \citep{RomanovaUstyugova2004}, instead of assuming a few separated ($~\le10$) accretion footpoints \citep{DonatiJardine2007}.

Therefore, the post-shock region has the vertical dimension $L$ significantly smaller than the other two dimensions. In Fig.~\ref{fig:accretion} we display a schematic view of the footpoint of an accretion stream and the post shock region. The resonance scattering effect on the X-ray spectrum should be different depending on whether we observe the emitting plasma from the top or from the side (inclination angle $\theta_{1}$ or $\theta_{2}$ respectively). The latter is the one that favors optical depth effects, while in the former case these effects could even be negligible. The different opacity effects observed for MP~Mus and TW~Hya could be explained assuming that for the two stars the angle $\theta$ of the accretion footpoints differ significantly. TW~Hya and MP~Mus are viewed under quite different inclinations $i$: TW~Hya is almost pole-on \citep[$i\la10\degr$, e.g.][]{QiHo2004}, while MP~Mus should have $i\ga60\degr$, considering that its optical spectrum indicates $v\sin i\sim13\,{\rm km\,s^{-1}}$ and that its rotational period is $\sim5-6$\,d \citep{BatalhaQuast1998}. Assuming that the footpoints of the accretion streams are located at high latitudes on both stars \citep[i.e. the magnetic dipole field is almost aligned with the rotational axis,][]{RomanovaUstyugova2004}, the orientation of TW~Hya should be similar to the top-view ($\theta_1$ in Fig.~\ref{fig:accretion}), while that of MP~Mus to the side view ($\theta_2$). We suggest that this important difference in viewing angle $\theta$ might explain the different optical depth obtained from the X-ray spectra of the two accreting stars. We cannot exclude that, although the inclination angles $i$ of two stars differ significantly, the two viewing angles $\theta$ of the post shock region are similar (e.g. if the accretion footpoint latitude is $\sim50-60\degr$). If it were the case, a different explanation would be required.

\subsection{Soft X-rays from shock-heated plasma: EMD}

The above results are consistent with the current prevailing view that accreting stars have two distinct plasma components: coronal plasma and shock-heated plasma, both contributing to different extents to the observed X-ray emission. To study the properties of these two plasma components it is necessary to identify and disentangle them with the help of both density and temperature diagnostics.

\begin{figure} 
\resizebox{\hsize}{!}{\includegraphics{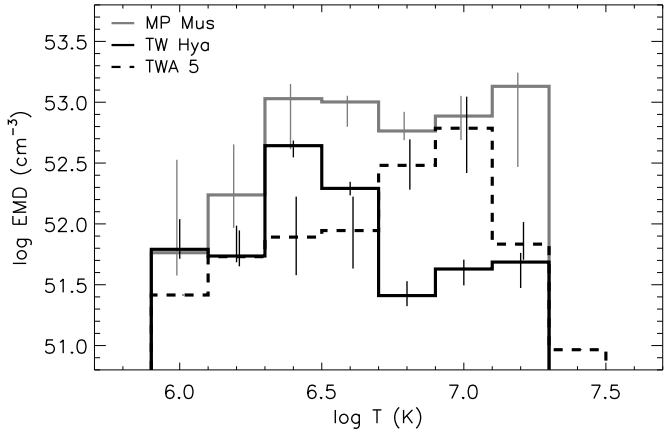}}
\caption{Emission measure distributions derived from the RGS line fluxes.}
\label{fig:allemd} 
\end{figure}

The {\it EMD} can also provide information on the nature of the X-ray emitting plasma. In Fig.~\ref{fig:allemd} we show the {\it EMD} of the two CTTS MP~Mus and TW~Hya. To understand their {\it EMD} in terms of the coronal and accretion-driven hot plasma components, we compared them with the {\it EMD} of TWA~5 \citep{ArgiroffiMaggio2005}, rebinned on the same $\log T$ grid used for the {\it EMD} of MP~Mus and TW~Hya.

The {\it EMD} reconstruction is based on the assumption that emission is optically thin. We know that this hypothesis is not true in the case of MP~Mus.  The lines affected are the resonance lines of \ion{O}{vii} and \ion{O}{viii} which constrain the cooler part of the {\it EMD}. Hence, it must be noted that the true {\it EMD} of MP~Mus at low temperatures ($\log T \la 6.5$) is likely to be slightly higher than that inferred by a factor of 2 or so.

Inspecting Fig.~\ref{fig:allemd} we found three main results:
\begin{enumerate}
\item MP~Mus and TW~Hya show a peak in their $EMD$ at low temperature ($\log T\sim6.4$); this peak is not present in the $EMD$ of TWA~5;
\item MP~Mus and TWA~5 have a strong $EMD$ peak at high temperature ($\log T\sim7.0-7.2$), while for TW~Hya that peak, even if marginally detected, is significantly lower compared to those of the other two stars;
\item the relative strength of cool and hot peak $EMD$ in MP~Mus and TW~Hya is significantly different.
 
\end{enumerate}

\begin{figure} 
\resizebox{\hsize}{!}{\includegraphics{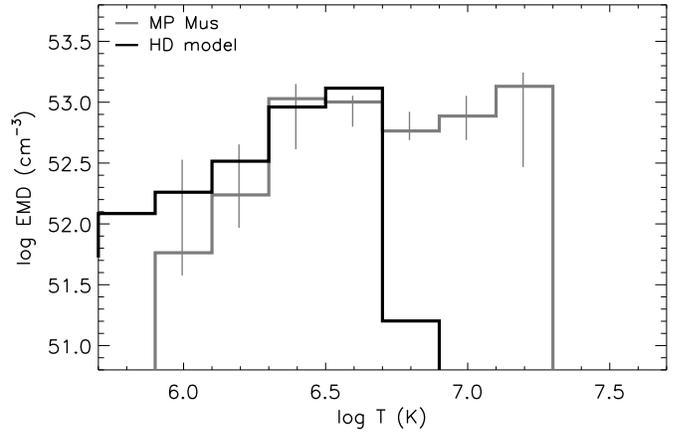}}
\caption{{\it EMD} of MP~Mus derived from the RGS line fluxes, and the {\it EMD} derived from hydrodynamical simulations of the accretion shock by \citet{SaccoArgiroffi2008}.}
\label{fig:emdandmodel} 
\end{figure}

\citet{SaccoArgiroffi2008} performed hydrodynamic simulations, tuned to the MP~Mus case, to study the shock formed by an accretion stream impacting on a chromosphere and its X-ray emission. In Fig.~\ref{fig:emdandmodel} we show the $EMD$ of MP~Mus together with the time-averaged $EMD$ of the simulation by \citet{SaccoArgiroffi2008}. The normalization of the model {\it EMD} was derived assuming that the soft X-ray emission is entirely produced by the shock-heated plasma component. This is reasonable, considering the \ion{O}{vii} line triplets: any contribution from coronal plasma, assuming that it has low density, cannot exceed 20\% \citep{ArgiroffiMaggio2007}. The important result is that the model {\it EMD} has a pronounced peak at $T\sim3$\,MK, which is exactly the feature that we observe in MP~Mus and TW~Hya, while no cool peak is present in the {\it EMD} of the WTTS TWA~5.

The cool peak that we found in the {\it EMD} of the two CTTS TW~Hya and MP~Mus explains, in terms of {\it EMD}, the soft flux excess observed in the X-ray spectra of CTTS in the XEST survey \citep{GuedelTelleschi2007,TelleschiGuedel2007}.  

These results suggest that the shape of the {\it EMD} in CTTS can be used to determine whether soft emission can be produced by accretion shocks, also in those cases in which the electron density provided by the \ion{O}{vii} triplet is compatible with coronal plasma \citep[i.e. the case of T~Tau,][]{GuedelSkinner2007}, or when the vital diagnostic intercombination and forbidden lines are of poor quality.

\subsection{Soft X-rays from shock-heated plasma: mass accretion rate}

\begin{table}[t]
\renewcommand{\baselinestretch}{1.3}
\caption{Mass accretion rate.}
\label{tab:accrate}
\normalsize
\begin{center}
\begin{tabular}{l@{\hspace{8mm}}c@{\hspace{1mm}}c@{\hspace{8mm}}c@{\hspace{1mm}}c}
\hline\hline
Star   & $\dot{M}$         & Ref. & $\dot{M}_{\rm X}$ & Ref. \\
\hline
MP~Mus & $3\times10^{-9}$  & (0)  & $8\times10^{-11}$ & (1) \\
TW~Hya & $2\times10^{-9}$  & (2)  & $2\times10^{-10}$ & (3) \\
BP~Tau & $2\times10^{-8}$  & (4)  & $9\times10^{-10}$ & (5) \\
\hline
\end{tabular}
\end{center}
\renewcommand{\baselinestretch}{1.0}
(0) this work; (1)~\citet{SaccoArgiroffi2008}; (2) \citet{AlencarBatalha2002}; (3)~\citet{GuntherSchmitt2007}; (4)~\citet{ArdilaBasri2000}; (5)~\citet{SchmittRobrade2005}.
\normalsize
\end{table}
\normalsize

The comparison between the {\it EMD} of MP~Mus with that of the WTTS TWA~5, and the results obtained from the hydrodynamical simulations performed by \citet{SaccoArgiroffi2008} shows that the soft X-ray emission of MP~Mus can be entirely explained in terms of the plasma heated in the accretion shock. The mass accretion rate provided by the simulations, $8\times10^{-11}\,{\rm M_{\odot}\,yr^{-1}}$, agrees with that derived on the basis of a simplified model by \citet{ArgiroffiMaggio2007}. However the $\dot{M}$ value derived from the H$\alpha$ line width exceeds that obtained from X-rays, $\dot{M}_{\rm X}$, by more than one order of magnitude. This discrepancy was noted also by \citet{Drake2005} and \citet{GuntherSchmitt2007} for TW~Hya. A similar situation arises for BP~Tau, for which mass accretion rates were derived from X-ray data by \citet{SchmittRobrade2005}. We list in Table~\ref{tab:accrate} mass accretion rates derived from X-ray data for these three CTTS, compared to those derived from other accretion indicators (i.e. UV or H$\alpha$). In all three cases the $\dot{M}_{\rm X}$ values are lower by a factor 10 (or even more) than the corresponding value derived from UV or optical data.

We note that the observed fluxes of the strongest lines were used to derive $\dot{M}_{\rm X}$, assuming optically thin emission. Since we found that the X-ray emission produced by the shock-heated plasma, in MP~Mus, is not optically thin, the accretion rate of MP~Mus derived from X-rays, \citep[$5-8\times10^{-11}\,{\rm M_{\odot}\,yr^{-1}},$ derived from the Ly$\alpha$ and resonance lines of \ion{O}{viii} and \ion{O}{vii},][]{ArgiroffiMaggio2007,SaccoArgiroffi2008}, is likely underestimated by a factor of 2 or so---much too little to reconcile the discrepancy between the $\dot{M}$ values.

The accretion flow is likely composed of several funnels, each isolated from the adjacent one because the strong magnetic field inhibits thermal conduction and mass motion perpendicular to its lines. Each funnel is characterized by a given density and infall velocity. Both the density and velocity determine the amount of observable X-rays produced by the accretion shock of each funnel. As explained below a distribution of density and velocity might explain the observed $\dot{M}$ discrepancy.

\citet{Drake2005} argued on the basis of accretion stream ram pressure and shock stand-off height that more dense streams would form smaller, less extended shocks buried too deeply in the photosphere to be observed from most inclination angles. X-rays would then be reprocessed to lower energy by the surrounding photospheric gas. The observed accretion-shocked plasma of MP~Mus should have a stand-off height of a $10^8-10^9$\,cm or so \citep[e.g.][]{ArgiroffiMaggio2007,SaccoArgiroffi2008}---sufficient to lie well above the photosphere and be visible with minimum absorption from the stellar atmosphere. It is possible that there are other streams of accretion carrying the majority of the inflowing mass flux that are much more dense and that do get obscured from the line-of-sight by absorption. X-ray emission from TW~Hya apparently does not fit in this scenario: it is produced by a very high-density plasma, but it suffers low absorption. To solve this issue, \citet{Drake2005} suggested that the plasma density of TW~Hya could be slightly overestimated, because of UV radiation and its influence on the He-like triplets.

Another effect might also contribute to the observed discrepancy. \citet{RomanovaUstyugova2004} and \citet{GregoryJardine2006} inferred that accreting material arrives at the base of the accretion stream with a large distribution of infall velocities. Therefore assuming that different funnels (or different part of the same funnel) may have different velocities, then each funnel will produce a post shock region with a temperature depending only on the relevant velocity. In such case, funnels with high velocity would produce hot post shock plasma, and, hence, X-ray emission, while funnels with reduced velocity might produce post-shock plasma of insufficient temperature to contribute to the X-ray emission.

\subsection{Coronal X-ray luminosity in CTTS}

The presence of coronal plasma for both TW~Hya and MP~Mus is indicated by their observed flaring activity and by the presence of a hot plasma component detected in their EPIC spectra, and inferred by the presence of emission lines produced by hot plasma \citep[$T\sim10$\,MK,][]{KastnerHuenemoerder2002,StelzerSchmitt2004,ArgiroffiMaggio2007}. Assuming that the cool peak in the {\it EMD} of MP~Mus and TW~Hya is entirely due to accretion, then their coronal component can be evaluated considering only the hottest part of their {\it EMD}. If we take into account only $\log T (K) \ge6.7$, the coronal X-ray luminosity of MP~Mus and TW~Hya are $1.4\times10^{30}$ and $1.4\times10^{29}\,{\rm erg\,s^{-1}}$, respectively, in the $0.5-8.0$\,keV band. The coronal luminosity of TW~Hya is significantly reduced if compared to its whole X-ray luminosity, $8.1\times10^{29}\,{\rm erg\,s^{-1}}$, while for MP~Mus the corona emits $\sim80$\% of its entire X-ray luminosity. With these new estimates for the coronal $L_{\rm X}$, MP~Mus and TW~Hya have $\log (L_{\rm X}/L_{\rm bol})=-3.4$, and $-3.8$ respectively. Both of these values, but especially that of TW~Hya, are significantly lower than previous estimates of their coronal to bolometric luminosities \citep{MamajekMeyer2002,KastnerZuckerman1997}, and place the two CTTS under the saturation level $\log (L_{\rm X}/L_{\rm bol})=-3$, as is usual for CTTS \citep[e.g][]{PreibischKim2005}. This suggests that, when the X-ray luminosity is to be used as activity indicator, it would be preferable to compute it excluding the plasma components at $T\le5$\,MK.

\section{Conclusions}

In this work we presented an analysis of the high-resolution X-ray spectra, obtained with {\it XMM-Newton}/RGS, of the two CTTS TW~Hya and MP~Mus.  For MP~Mus we detected significant resonance scattering in the \ion{O}{viii} Ly$\alpha$ and \ion{O}{vii} resonance lines, which are produced by the high-density plasma at temperatures of a few MK. No  resonance scattering was detected in the spectrum of TW~Hya.  This result strongly supports the hypothesis that this plasma is formed by the shock at the base of the accretion column which is likely viewed obliquely.  The different optical depths observed for TW~Hya and MP~Mus could be explained in terms of different viewing angle of their accretion shocks.

We also derived the {\it EMD} for TW~Hya and MP~Mus, finding that they both show a peak at $T\sim3-4$\,MK, in addition to the hot peak at $10-20$\,MK typical of coronal plasma on magnetically active stars. The cool peak is perfectly described by plasma heated in an accretion shock \citep{SaccoArgiroffi2008}. The same peak is not present in the plasma of the non-accreting young star TWA~5. The identification of this {\it EMD} peak as due to shock-heated plasma allows us to assess the characteristics of the true coronal plasma and its luminosity. In particular, the coronal X-ray luminosity of TW~Hya is less than 20\% of its whole X-ray luminosity in the $0.5-8.0$\,keV band.

Soft X-ray emission can be used to compute the mass accretion rate. We compared the mass accretion rates derived from X-ray data, $\dot{M}_{\rm X}$, with those obtained from UV and optical indicators, $\dot{M}$, finding that $\dot{M}_{\rm X}$ is underestimated by, at least, a factor 10. Two possible explanations for this are that some accretion streams are most dense and form shocks that are located too deep in the stellar atmosphere to be observed, and that some plasma in accretion streams might not attain the full free-fall velocity required to form X-ray emitting shocks.

\begin{acknowledgements}

The authors thank the referee, M.~G{\"u}del, for comments that improved the paper. CA, AM, GP, SS, and BS acknowledge partial support for this work from contract ASI-INAF I/023/05/0 and from the Ministero dell'Universit\`a e della Ricerca. JJD was supported by NASA contract NAS8-39073 to the {\em Chandra X-ray Center} during the course of this research, and thanks the director, H.~Tananbaum, and CXC science team for advice and support. JLS acknowledges financial support by the projects S-0505/ESP-0237 (ASTROCAM) of the Comunidad Autonoma de Madrid and AYA2008-06423-C03-03 of the Spanish Ministerio de Ciencia y Tecnolog\'{\i}a. Based on observations obtained with {\it XMM-Newton}, an ESA science mission with instruments and contributions directly funded by ESA Member States and NASA.

\end{acknowledgements}

\bibliographystyle{aa} 
\bibliography{mpmus}

\end{document}